 \documentclass[conference]{IEEEtran}
 \usepackage{graphicx}
 \graphicspath{{images/}}

 \author{
\IEEEauthorblockN{Sarah C. Helble, Alexander J. Gartner, Jennifer A. McKneely}
 \IEEEauthorblockA{Johns Hopkins University Applied Physics Laboratory\\
 firstname.lastname@jhuapl.edu}
 }

\begin{document}

\title{Increasing the Security of Weak Passwords: the SPARTAN Interface}

\maketitle
\begin{abstract}

  Password authentication suffers from the well-known tradeoff between security
  and usability. Secure passwords are difficult for users to remember, and
  memorable passwords are often easy to guess. 
  SPARse Two-dimensional AuthenticatioN (SPARTAN) allows users
  to input their textual passwords in a two-dimensional grid instead of a linear
  textbox. This interface enables relatively short passwords to 
  have a higher calculated level of 
  security due to the need for an attacker to determine both the text of the 
  password and the location of each character in the grid. 
  We created a SPARTAN prototype and conducted a preliminary user study to 
  evaluate the actual usability and security of the
  SPARTAN interface compared to the linear password entry interface.
  We find that while user-created SPARTAN passwords tend to be shorter than 
  their linear counterparts, the calculated security of user-created SPARTAN 
  passwords
  is higher than that of user-created linear passwords.
  We also asked participants to complete a survey on the usability of the 
  SPARTAN interface and identified some areas of improvement, while prototype
  interaction provided some
  evidence of users becoming more familiar with SPARTAN over time.
  Finally, we performed an investigation into password-cracking tools, and assert
  that SPARTAN passwords require more resources to crack than their linear
  counterparts.
  These findings suggest that SPARTAN is a promising alternative to linear 
  passwords from a security standpoint. Usability of the interface and 
  memorability of SPARTAN passwords is an interesting research question
  and should be further investigated in future work.
  
\end{abstract}

\section{Introduction}
\label{sec:introduction}

Password authentication is in many ways both a blessing and a curse for modern 
computer systems and applications and their users. Although many researchers and
users would enjoy the riddance of password authentication in their daily lives, 
the relative ease of deployment and usability of password authentication shows 
that passwords are here to stay for the foreseeable future 
\cite{QUEST:PW_STAY}\cite{PW_PERSIST}.
However, when given the freedom to choose their own passwords, many users will 
opt to create passwords that are easy for an adversary to guess or
easy for a password recovery tool to crack \cite{LORI:TED}. 
Users can be trained or given tools and guidelines to aid in the creation of 
more secure passwords. However, studies have found that most users 
will actively try to circumvent these requirements in order to create a password
they are able to remember \cite{SOUPS:DAILY_AUTH}. 

This paper presents
SPARse Two-dimensional AuthenticatioN (SPARTAN), a password entry 
interface that attempts to bridge the gap between security and usability. 
SPARTAN allows a user to input each character of his password anywhere in a 
two-dimensional grid
instead of a linear textbox. Even with sparsely-populated grids, SPARTAN 
passwords increase entropy over linear passwords due to the variability of the 
location of each character. In this way, users can have 
simpler passwords while achieving a higher overall password 
security.

As prior work with SPARTAN has been largely theorectical, we 
conducted a preliminary study to test the usability and 
security of the SPARTAN interface with unfamiliar users. 
Our study was 
especially concerned with the time required for users to create a password 
using our interface, and the resulting security of the password created both
in terms of calculated security (i.e., entropy) and security against a stolen
password file attack. 
The study reached a total of 100 participants, 48 of which interacted with the 
SPARTAN prototype. The remaining 52 participants formed a control group and were
asked to perform tasks on a linear password interface. We found that while 
passwords created with the SPARTAN interface were, on average, shorter and 
comprising fewer character sets than passwords created with the linear 
interface, SPARTAN users did choose a variety of starting positions and password
formations, which resulted in a higher average security for SPARTAN passwords
compared to their linear conterparts. 
Feedback on the SPARTAN
prototype mixed excitement and criticism, and the current prototype needs
further evaluation and focused usability development before use in an 
operational environment. 

The next section goes into more detail about SPARTAN, while Section 
\ref{sec:related} discusses related work in usable password authentication. 
Section \ref{sec:method} explains the methodology of our user study and the
functionality of the SPARTAN prototype interface created for this study. 
Our findings are presented in Sections \ref{sec:results} and \ref{sec:cracking},
while Section \ref{sec:discussion} gives a more academic discussion of the 
calculated security of user-created SPARTAN passwords. 
Finally, Section \ref{sec:conclusion} explains future 
work that we wish to perform to further evaluate SPARTAN and concludes.

\section{SPARTAN}

\begin{figure}
  \centering
  \includegraphics[scale=.35]{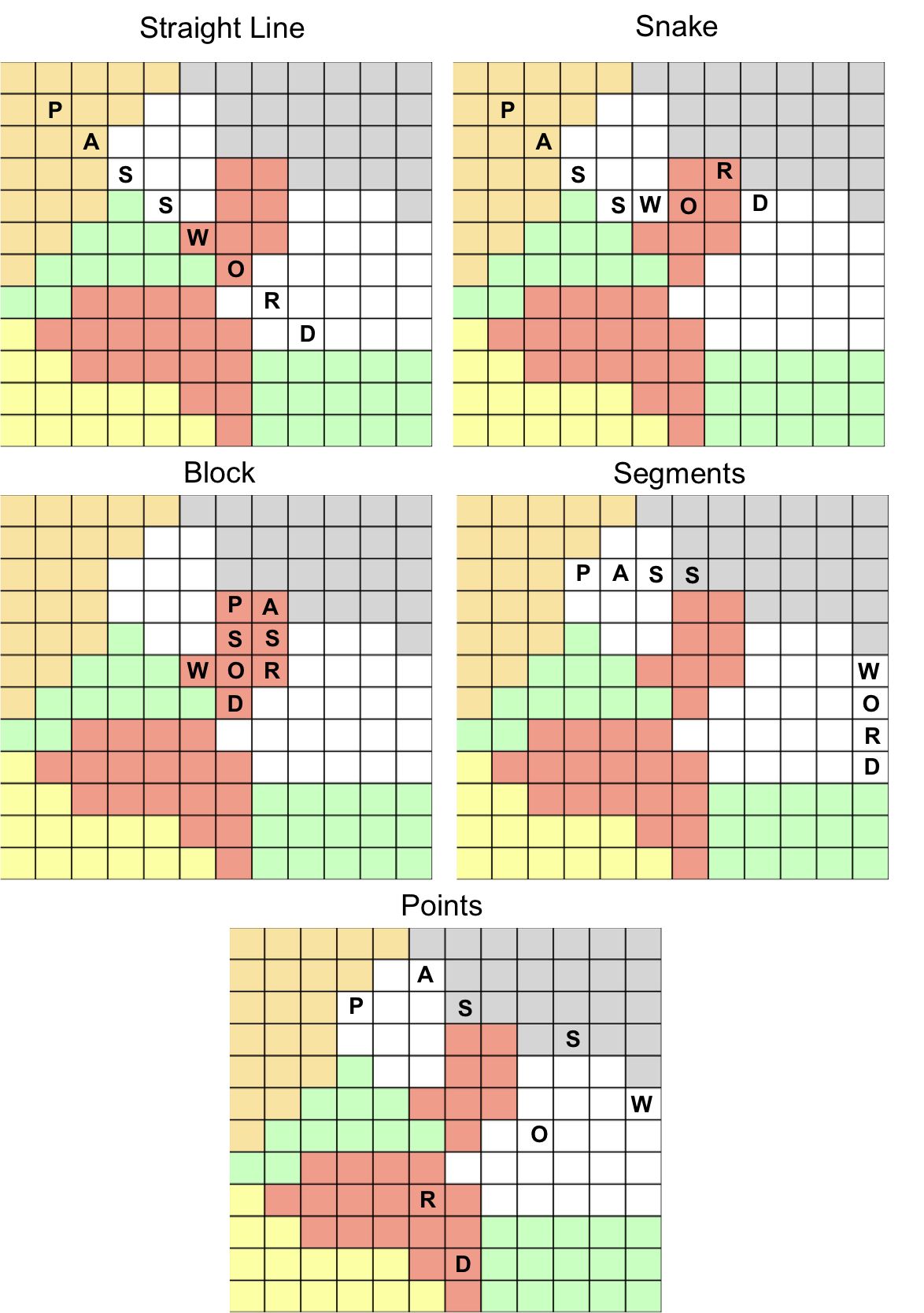}
  \caption{Different categories of SPARTAN passwords. A Straight Line password 
    continues in a single straight line through the grid. A Snake password 
    changes direction as it flows through the grid, but always continues into 
    an adjacent cell. A Block password fills a square or rectangle pattern
    in the grid, likely following the coloring scheme. A Segments password is a 
    Straight Line password composed of multiple segments, and a Points 
    password is made up of multiple disjoint points throughout the grid. 
}
  \label{fig:types}
\end{figure}

The SPARTAN interface allows the user to input their password
anywhere in a two-dimensional grid. The user's SPARTAN password is then the text
of the
password itself and the location of each character in the grid. Figure 
\ref{fig:types} gives examples of various types of SPARTAN passwords users 
could create. As its name implies, SPARTAN grids are sparsely populated, that
is, the whole grid does not need to be filled with characters in order to see a 
security benefit over linear passwords. 
Table \ref{tab:space} compares the 
password space for 8-character linear passwords and SPARTAN passwords.

\begin{table}
\normalsize
 \begin{centering}
    \begin{tabular}{| l | l |}
      \hline
      Password type      & Password Space\\\hline
      Linear             & 41 bits \\
      SPARTAN 12x12 grid & 98 bits \\
      \hline

    \end{tabular}
    \caption{Password Space of random linear passwords of 8 characters and
    random SPARTAN passwords of the same length, using an alphabet of 36 
    characters. For linear passwords, the equation used is $log_2(36^8)$. For
    SPARTAN passwords, this is $log((36^8)*( P(12*12, 8)))$}
  \label{tab:space}
  \end{centering}
\end{table}

To use SPARTAN, the user first selects a starting location. In deployment 
scenarios, this cell could be randomly chosen or 
otherwise constrained to help the user select a unique placement.
There is no requirement for the characters of a password to be placed in 
adjacent cells. In fact, SPARTAN passwords are most secure when the user 
chooses distinct points throughout the grid.
However, to aid ease of input, our implementations 
give the user the option to select 
a direction for their password. Direction can be changed during password entry; 
the prototype will automatically place typed characters into adjacent
cells in the selected direction. Order of character 
entry is not collected as part of the password, but SPARTAN does not preclude 
this.

The grid used for SPARTAN passwords can be as large or as small as necessary
to balance desired usability and security. Optimal grid size is an 
interesting area of future work we wish to pursue in a follow-on user study. 
In our prototypes, the squares of the SPARTAN grid have been colored in 
an abstract pattern in an attempt to limit the prevalance of hotspots in the 
grid, and to help users remember the placement of their passwords. 
This colorization is merely a visual aid provided by the interface; it is not 
chosen by the user as part of their password.
Colorization would be different across accounts, so it could also aid users
in remembering multiple different SPARTAN passwords.

The SPARTAN grid also employs a wrapping mechanism so that the user does not 
have to estimate spacing before creating his password. In our
implementations this mechanism can result in the overwriting of characters. 
Future implementations could fix this by allowing
multiple characters to reside in a single cell. 
The encoding of the SPARTAN password is an implementation decision, and can be
any method which preserves each character of the password and its location. 
One method could be to encode the password as follows:

\begin{verbatim}
          23P24a25s26s36w46o56r57d
\end{verbatim}

Where the initial '23' is location of the 'P' in the grid, '24' is the location
of the 'a', and so forth. The encoded password can then be hashed and
stored in the same way as linear passwords.

\subsection{Deployment Options}

The SPARTAN interface can be used anywhere that linear passwords are used, 
including as part of a two-factor authentication scheme. We have developed a 
JavaScript SPARTAN interface, which can be incorporated into any existing 
web-based login page.
We have also created a SPARTAN user script which can be deployed client-side, 
with no changes made to the authentication server. In this use case, the user
installs a SPARTAN user script in their browser.
The user can then enter a backslash ('\textbackslash') or another 
user-defined character into any password field and the SPARTAN grid will appear
on top of the webpage. The user can then enter his SPARTAN password in the grid,
and the hashed version of this password will be automatically placed in the 
website's linear password field.

\section{Related Work}
\label{sec:related}

The topic of usable passwords has been a contentious subject of research
for many in the field of privacy and security throughout the years. 
Among the participants in a study by Mare et al., passwords 
were the most commonly disliked form of authentication, and had a relatively 
high failure rate among users \cite{SOUPS:DAILY_AUTH}. Many other experts and 
studies find that current password practices are burdensome to users 
\cite{LORI:TED}. However, due to 
the deployability of passwords, among other factors, passwords are here to stay 
\cite{QUEST:PW_STAY}\cite{PW_PERSIST}.

Password managers can help solve the common user problem of having too many 
passwords to remember; however, a
study by Ion et al. found that many users are distrustful of 
password managers and do not find them to be user-friendly 
\cite{ION:EXPERT_VS_NON}. Furthermore, many organizations do not feel they can
trust third-party password managers, and restrict their employees from using 
them for work accounts \cite{SOUPS:DAILY_AUTH}.

A study by Shay et al. evaluated the tradeoff between security and usability in 
user-created passwords made under different length and complexity 
requirements \cite{SHAY:COMPLEX_TRADEOFF}. 
Some studies have suggested the use of randomly-chosen dictionary words for 
passwords. A study by Cranor et al. found that computer-generated passphrases
made up of randomly-chosen dictionary words
are not any more memorable than computer-generated short passwords, and the 
passphrases take longer to type (and re-type if the user makes an error) 
\cite{LORI:TED}. 
In their work with Persuasive Text Passwords (PTP), 
Forget et al. combined user- and computer-generated passwords.
PTP increases the security of user-created passwords by placing random 
characters
at random positions in the password text. Their study found
that PTP is a promising method both in terms of usability and security, but also
discovered a limit where users would create weaker initial 
passwords if PTP was too aggressive in its transformations \cite{FORGET_PTP}.

Mnemonic passwords, where users are instructed to make a password by choosing a 
character for each word of a selected phrase, have also been an interesting 
point of research. Kuo et al. found that user-created mnemonic 
passwords are often based on well-known phrases, such as song lyrics, and
they could be vulnerable to a specially-crafted dictionary attack \cite{MNEMON}.
A study by Forget et al. found that these password schemes may be difficult for
users to understand and are easy to misuse \cite{FORGET:MNEMON}.

Graphical passwords are another variation on normal password authentication that
is currently in use on various mobile and desktop platforms. During their 
study using the graphical password system PassTiles, Strobert and Biddle
showed that graphical passwords are easier for users to remember than 
generated textual passwords \cite{passtiles}. However,
they can also be prone to predictability, and can even fall victim to dictionary
attacks \cite{graphical-twelve}\cite{GRAPH} or the selection of 'hot-spots' in 
the chosen image \cite{HOTSPOTS}. 
Tao and Adams observed that users create relatively long graphical passwords 
with their 
Pass-Go interface, wherein users pick and connect intersections on a grid 
rather than the conventional selection of grid cells. Their research also shows
a propensity for users to start in predictable locations in the grid and to 
draw familar shapes \cite{passgo}.
Chiasson et al. developed Persuasive Cued Click-Points 
(PCCP) to aid users in selecting more random locations in a graphical password,
with significant success \cite{PCCP}. However, many forms of graphical 
passwords also fall 
short when compared to textual passwords because they must allow a degree of 
tolerance in user input. Without this, users are often unable to correctly 
enter their graphical passwords \cite{TOLERANCE}. 

Safdar and Hassan suggest the completion of a full
two-dimensional grid as passwords in secure environments 
\cite{2D_1}. In their research, they acknowledge that this method should only be
used in extreme 
situations where security is paramount, because the passwords are too complex 
and lengthy for normal users to remember.

Perhaps the most similar work to the research described in this paper is 
Saharkar and Dhopte's scheme for the combination of graphical and alphanumeric
passwords. In their model, they suggest that users should 
choose points of interest on an image, and have the option of
adding a textual password to each point upon password entry 
\cite{GRAPH_ALPHA_2}. As of this writing,
there is no published research evaluating the security or usability of this 
method. 

Our study is unique to all previous research in that it evaluates the 
combination of aspects of graphical and linear passwords. Although SPARTAN uses
a two-dimensional grid, it is different from Safdar and Hassan's work in 
that its use does not require users to complete the entire grid 
space. In this way, our study gives unique insight into the
creation of passwords and is the first to evaluate the SPARTAN method of 
authentication.

\section{Method}
\label{sec:method}

We conducted a Institutional Review Board (IRB)-approved human 
factors study with adult employees of our organization in
order to evaluate the usability of our SPARTAN prototype and differences between
user-created SPARTAN and linear passwords. During the course of this study, we 
were especially concerned with identifying usability limitations 
in our SPARTAN prototype, identifying any potential patterns in participant's 
use of the prototype and comparing passwords created with a linear password 
interface with those created with the SPARTAN interface.

\begin{figure}
  \centering
  \includegraphics[scale=.7]{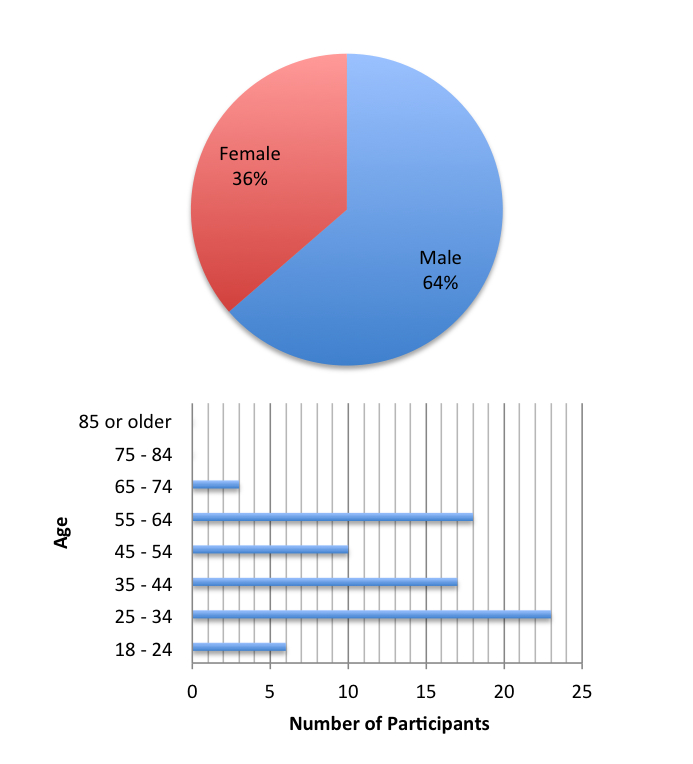}
  \caption{Participant-reported age and gender demographics. Notice that not 
    all participants completed the survey phase of the study, and they were
    not required to answer questions to complete their participation. N=77}
  \label{fig:demographs}
\end{figure}

\begin{figure}
  \centering
  \includegraphics[scale=.6]{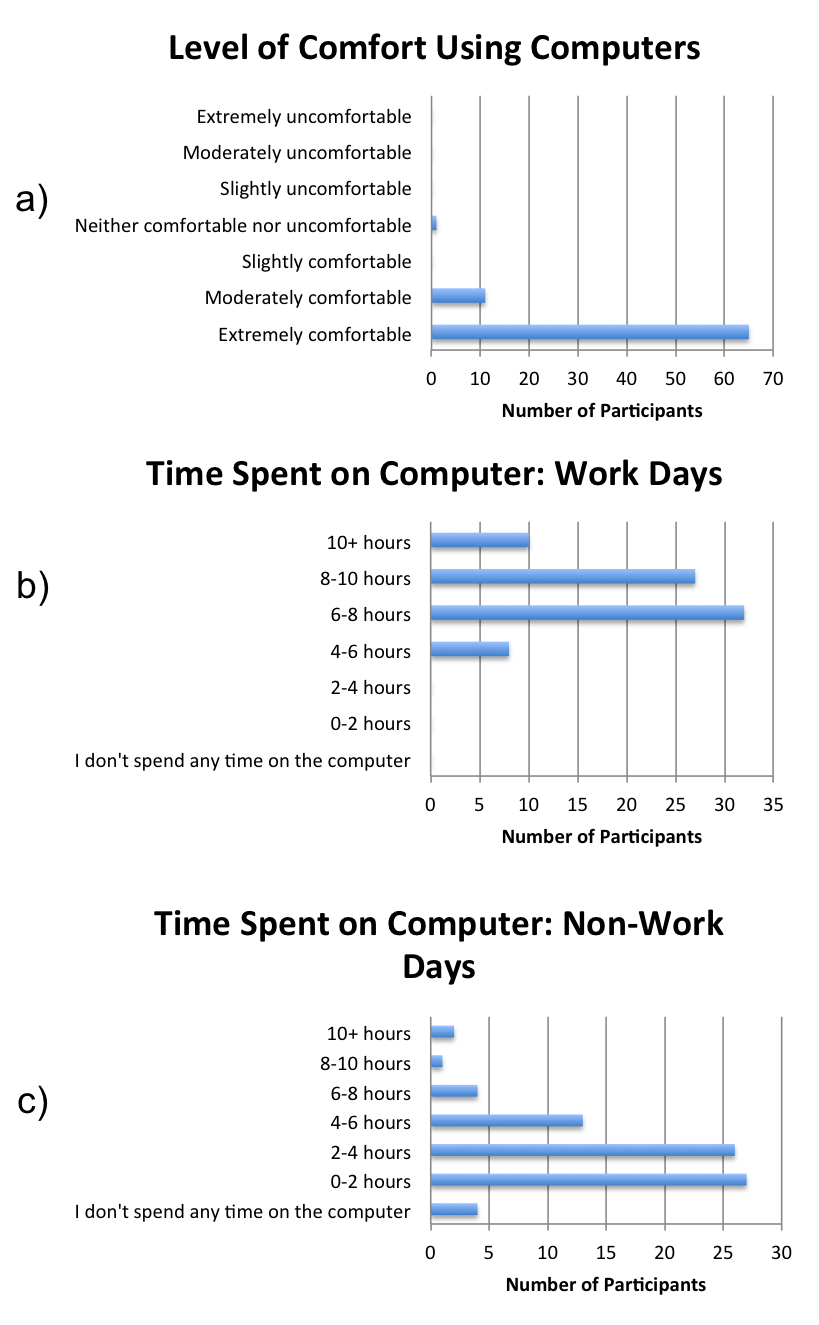}
  \caption{Participant-reported computer usage: a) shows participants' 
responses to the question ``How comfortable are you with using computers?'' b) 
depicts responses to 
``How much time on average do you spend on a computer on work days?'', and c) 
shows the distribution of answers to the question 
``How much time on average do you spend on a computer on non-work days?''.
 N=77}
  \label{fig:comp_use}
\end{figure}

Participants were recruited via email and could participate at their leisure 
from the comfort of their 
individual workstations. Figures \ref{fig:demographs} and 
\ref{fig:comp_use} show demographic information of participants and their 
reported computer usage. The study took place over the course of three weeks
from late August to early September 2016.

\begin{figure}
  \centering
  \includegraphics[scale=.6]{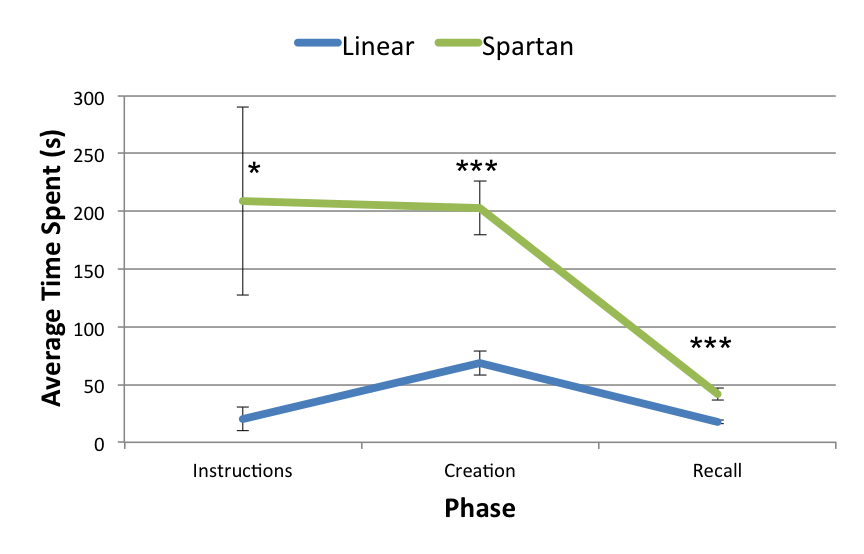}
  \label{fig:phase_time}
  \caption{Average time spent by participants in each phase. 
   ( $*p<0.05,**p<0.01,***p<0.001$ 2-tailed t-test ) }
\end{figure}

\subsection{Procedure}
\label{sec:procedure}

Participants were randomly assigned to the control group (linear passwords) or 
the experimental group (SPARTAN passwords). Each group went through four phases:
instruction, creation, survey, and recall. The study had to be completed 
in one session; users were not able to return to complete their participation. 

During instruction, the experimental group was given
minimal instructions on SPARTAN and an interactive demo of its use.
In the creation phase, members of each group created a password.
For both SPARTAN and linear participants, the only password
requirement was that it comprise at least eight characters. 
All passwords were masked during creation. 
In addition to a password hash, we collected a converted version of the 
passwords created which conveyed the character set used for each character
of the password (uppercase, lowercase, numerical, or special character). 
This data allowed us to determine the password length and number of 
character sets used as well as the placement of SPARTAN passwords 
in the grid.

After creating their passwords, all participants completed a survey for 
demographic information and password practices. SPARTAN 
participants were asked additional questions about the SPARTAN interface.
Participants were then asked to recall the password
they created at the beginning of the study, with the opportunity to retry as many
times as desired. 

Some participants did not complete all phases of the study; user retention is
discussed further in Section \ref{sec:retention}.
Of the participants that reached the recall phase, the median 
total time spent on the study for members of the control group was 3.8 minutes, 
whereas in the experimental group the median time was 13.7 minutes. Figure 
\ref{fig:phase_time} shows the average time spent in each phase for all
participants. 

\subsection{SPARTAN Prototype}
\label{sec:prototype}

The SPARTAN prototype used in this study is a 12x12 grid, with 
the block colorization seeded by the username used.
When entering a password, the user was expected to 
first choose a starting location by clicking on a cell in the
grid, then choose a typing direction for their password by clicking on an arrow
or pressing an arrow on their keyboard. In order to de-clutter the interface, 
we opted for arrows superimposed on the grid itself instead of a separate directional pad (d-pad).
As a goal of this study was to gather data on where users would independently 
place their passwords in the SPARTAN grid, we refrained from giving a default
grid location and direction. 

\begin{figure}
  \centering
  \includegraphics[scale=.43]{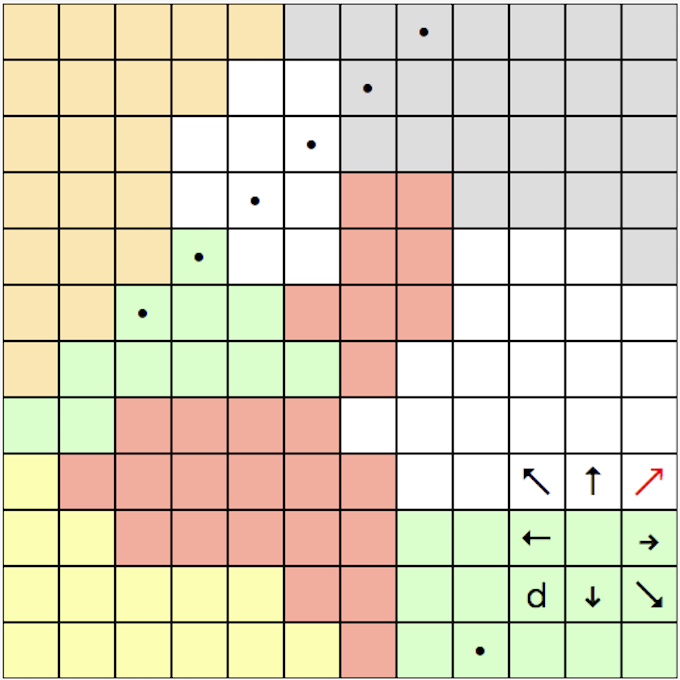}
  \caption{SPARTAN prototype used in the study, with overlaid arrows, 
    grid colorization, text wrapping, and mobile-style masking mechanisms.}
  \label{fig:prototype}

\end{figure}

We also implemented a mobile-inspired masking mechanism, 
where characters in adjacent cells remain visible to the user until the cursor 
is moved to a distant cell. Characters in each cell could be
unmasked by hovering over the desired cell. A screenshot
of the developed prototype displaying these features is shown in Figure 
\ref{fig:prototype}.

Some of the design decisions made for the use of this study 
were found to 
have adverse affects on the usability of the prototype. These concerns are 
explained further in Section \ref{sec:usability}.

Passwords created with the SPARTAN prototype were collected by 
filling empty cells with spaces in order to effectively  
preserve location data and concatenating each row to the preceding rows. The
result of the concatenation was then hashed to compare against re-entered 
passwords.

\section{Results}
\label{sec:results}

\subsection{Password Practices}
During the survey phase, we asked all participants about their password 
practices. When asked how often they forget a password to one of their accounts,
49\% of participants said ``Once or twice a month'' or more often (N=77). When 
asked how difficult it is to create a password that meets password 
requirements, 55\% of participants reported ``slightly difficult'' or higher 
(N=77). When asked if they had any other comments, suggestions, or opinions on 
password practices or other topics, there was a mix of beliefs and security
knowledge among our participants. Some indicated that they are unsure of what
constitutes a secure password. Others stated that they use a password manager
to manage all of their passwords across different accounts, while one
stated that they write down their passwords in order to keep track of them.
Some participants indicated that password complexity requirements is reaching a 
level such that it is difficult to develop a scheme to keep track of the 
different requirements and passwords
across sites, while several voiced their opinions that longer passwords, 
or passphrases, offer more of a benefit against password crackers than increased
complexity requirements. A couple also cited research that 
suggests enterprises should not mandate
periodic password changes. Another participant mentioned that they wish 
biometrics, such as fingerprint scanners, could be adopted more universally. 

In general, our participants'
responses indicated that they are burdened by password creation, with
many sites enforcing disparate complexity requirements. This is in line with 
other research and expert opinions in the field 
\cite{LORI:TED}\cite{LORI-PW_CHANGE}\cite{SOUPS:DAILY_AUTH}. Many participants
demonstrated an affinity for longer length requirements in place of 
complexity requirements. This is also the opinion supported by the NIST Draft
Digital Authentication Guideline \cite{NIST_DRAFT-DAG_2016}, and is the idea 
behind the SPARTAN interface, to decrease complexity requirements while 
increasing security.

\subsection{SPARTAN Usability}
\label{sec:spartan_analysis}

In this section we evaluate SPARTAN in terms of the time required to use
the SPARTAN interface, the rate of recall of SPARTAN passwords, and the 
reported usability and acceptance of SPARTAN among 
participants.

\subsubsection{Time Considerations}
\label{sec:time}

As shown in Figure \ref{fig:phase_time}, the average time for participants to 
create a SPARTAN password was three times that of their linear counterparts.
In addition, experimental group participants took 
approximately twice as long as their control group counterparts to recall
their passwords. Although this is a significant increase in time required for 
creation and recall, we find these results promising as the factor of increase 
decreased as participants became more familiar with the SPARTAN interface. 
Future work should further evaluate the time required for users to login using 
SPARTAN after they've gained familiarity with the interface over time.

\subsubsection{Recall Rate}
\label{sec:recall}
In total, 31 SPARTAN and 41 linear participants reached the recall phase
of the study. Of these SPARTAN participants,
three did not attempt to recall their password. Both the control group and the 
experimental group had one participant who was unable to recall his 
password after several attempts. Of the participants that did sucessfully recall
their passwords, two in the control group and three in the experimental group
required multiple attempts. In both groups, the average number of attempts 
among all successful participants was 1.1. All participants in the study only 
had to remember their passwords for a couple of minutes while they completed the
survey; however, it is promising that SPARTAN group participants displayed a 
recall rate similar to their linear counterparts.
Future work should evaluate the recall rate of SPARTAN passwords over a longer
period of time, and when multiple passwords need to be remembered across 
different accounts.

\subsubsection{User Reaction}
\label{sec:usability}

\begin{figure*}
  \centering
  \includegraphics[scale=.65]{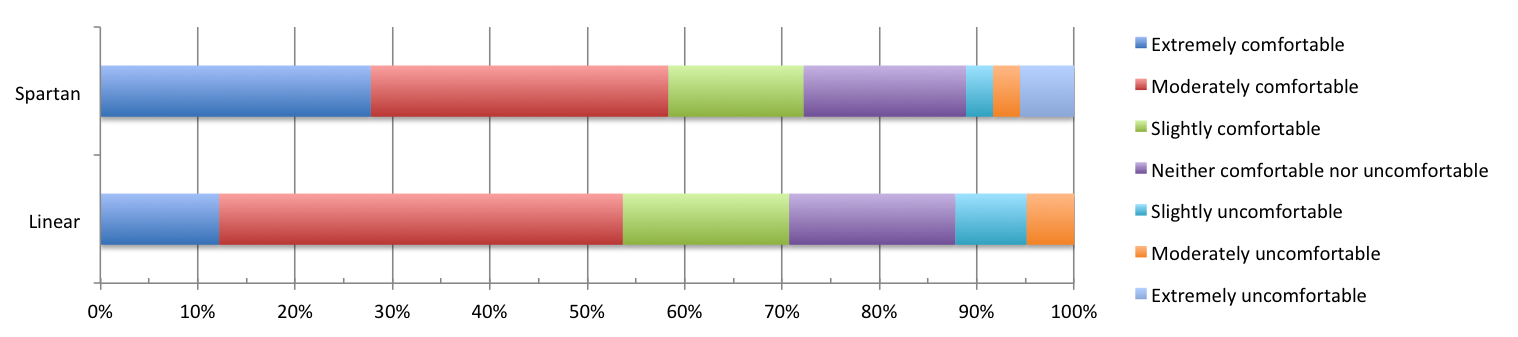}
  \caption{Participants' reported comfort with the password they created during 
the study. SPARTAN participants N=36, linear participants N=41. 
A greater number of SPARTAN participants than linear participants expressed 
extreme comfort with the password they created for their study, while an almost 
equivalent proportion of participants in each group expressed some positive 
level of comfort with their password.}
  \label{fig:comfort}
\end{figure*}

We asked members of both groups ``How comfortable are you with the security of 
the password you created during this study?''. A larger proportion of SPARTAN
participants reported being ``extremely,'' 
comfortable with the security of their password (28\%)
than linear participants (12\%). Approximately an equivalent proportion of 
SPARTAN and linear participants expressed some level of comfort with their 
created password (72\% of SPARTAN, 71\% of linear, N=36 and 41, respectively).
Please refer to Figure \ref{fig:comfort} 
for this data. In addition, a majority (78\%) of SPARTAN participants agreed 
with the 
statement ``I think SPARTAN passwords are more secure than traditional 
passwords,'' while a smaller majority (56\%) disagreed with the statement ``I 
think SPARTAN passwords are more memorable than traditional passwords'' (N=36). 

We also asked SPARTAN participants if they would prefer to use SPARTAN or 
traditional passwords in order to log in to their bank accounts. 37\% of users
said they would prefer SPARTAN while 63\% reported they would prefer linear 
(N=35).
The reasons behind these answers varied. The majority of participants in favor 
of SPARTAN cited improved security as the reason for their preference. 
Some even mentioned that they would prefer security over ease-of-use for 
important credentials. A few also mentioned that they liked that they 
could create a 
unique pattern that would be impossible for others to guess. Of those in favor
of linear passwords, a majority cited poor prototype usability as the reason 
behind their preference. 

One of the most prevalent problems with usability was caused by our decision to 
superimpose arrows on the grid instead of featuring a separate d-pad. With this
implementation, in order to move around the grid cell by cell, users could 
either click a cell or use the directional arrows on the keyboard to navigate. 
However, if a cell contained an arrow, one click would select the
arrow, and a second click would select the cell. This was understandably 
confusing to users unfamiliar with the mechanics. 
Our decision to eliminate default selections also resulted in 
poor usability. As noted in Section
\ref{sec:prototype}, in order to decrease bias, we did not implement a default
starting position or direction in the SPARTAN interface. This resulted in a
decrease in usability which would not be present in a deployed system. In 
Section \ref{sec:conclusion}, we propose future work to improve the usability
of SPARTAN. 

The next most frequent response in favor of linear passwords was a feeling that 
SPARTAN passwords would be harder to remember. As discussed in Section 
\ref{sec:recall}, we did not see a significant decrease in recall rate for 
SPARTAN passwords. However, our study only required a short amount of time 
between password creation and recall; future work should evaluate the recall 
rate of SPARTAN passwords over a longer period of time. A few participants 
indicated their preference was due to password managers not yet supporting 
SPARTAN passwords.

\subsection{SPARTAN Passwords}

\begin{figure}
  \centering
  \includegraphics[scale=.75]{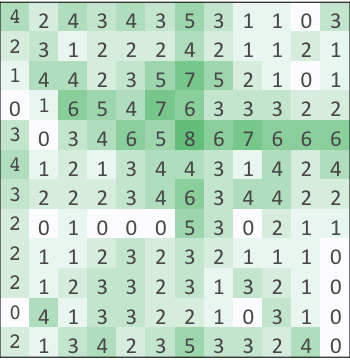}
  \caption{Heatmap of SPARTAN passwords created by participants. N=38 }
  \label{fig:heatmap}
\end{figure}

The increased security of SPARTAN passwords over linear passwords depends on the
variability of their placement in the grid. If all users choose the same 
location or set of locations, the security of SPARTAN passwords would decrease
and SPARTAN would not provide any additional security over linear passwords.
The data gathered during our study shows that users did choose variable 
password locations and strategies throughout the grid, and increased the 
security of their passwords by doing so. Figure \ref{fig:heatmap} shows the 
distribution of filled cells in the grid for all SPARTAN passwords created by 
users. 

\begin{figure}
  \centering
  \includegraphics[scale=.5]{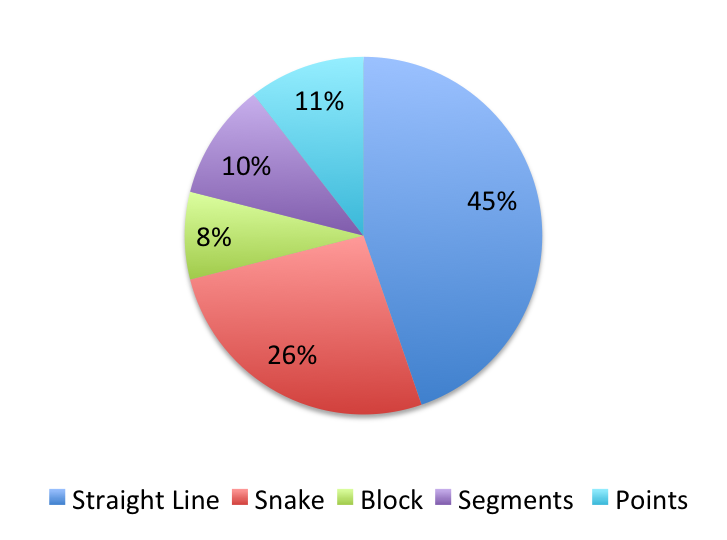}
  \caption{Categorization of participants' SPARTAN passwords created during the 
study. N=38}
  \label{fig:type_percents}
\end{figure}

In order to evaluate the types of passwords created by users, we divided them 
into five different categories: Straight Line, Snake, Block, Segments, and 
Points. Straight Line passwords are those that continued in a straight line 
through the grid. These could be horizontal, vertical, or diagonal. Snake 
passwords are those that comprise adjacent cells but can change 
direction any number of times as they travel through the grid. Block passwords 
were 
usually created by a user filling in all of the cells of a certain color block; 
Segment passwords occurred when the user divided their password into a number
of Straight Line segments and placed those segments in disjoint cells of the 
grid. Point 
passwords are those where the user chose to place each character of their 
passwords in disjoint cells across the grid. An example of each type of password
is given in Figure
\ref{fig:types}. The number of each type of password created during this study
is given in Figure \ref{fig:type_percents}. While the largest proportion of 
participants created Straight Line passwords, we were surprised to find that 
no single type of password was created by a majority of users. In addition, of
the 17 Straight Line passwords created in this study, eleven were horizontal, 
while two were vertical and four diagonal. Snake 
passwords had an average of 2.7 direction changes, and Segment passwords 
comprised an
average of 2.3 segments. These are promising metrics as they show that 
users will make a variety of SPARTAN passwords that cannot be easily predicted.

\subsection{Limitations}
\label{sec:study_factors}

There are a number of factors that could have affected the outcome of our study.
Members of the experimental group took much longer to complete their 
participation than members of the control group. This could have led to fatigue
in the experimental group.

As mentioned in Section \ref{sec:prototype}, in order to refrain from imposing
bias on the location and direction of participants' SPARTAN passwords, we did 
not give a default location or direction in the grid. This led to some usability
concerns among participants.

Finally, participants may have created stronger passwords than they normally 
employ because they knew the passwords were being monitored for a study. This is
especially true in the control group, as some participants indicated
a belief that the results of the study would be used to influence
password complexity requirements for our organization as a whole.

\subsubsection{Retention}
\label{sec:retention}

Although we recruited 100 participants from throughout our organization, not all
participants finished all phases of the study. 
Two participants each from the control and experimental groups quit 
after failing to create a password due to being unable to meet requirements 
and/or re-enter the password correctly. Since the same number of participants 
in each group did this, terminating at this time 
should not be attributed to difficulty with the SPARTAN interface.

\begin{figure}
  \centering
  \includegraphics[scale=.5]{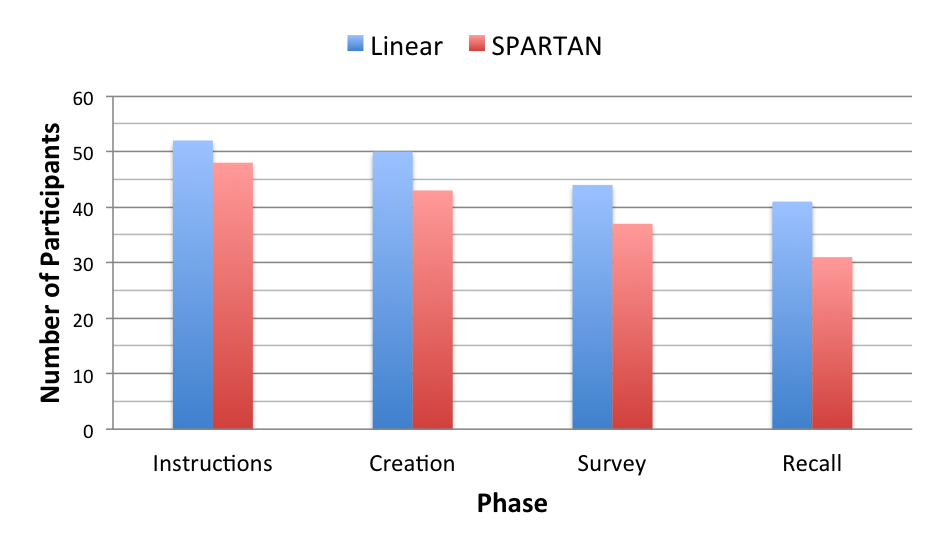}
  \caption{Number of participants in each phase of study.}
  \label{fig:retention}
\end{figure}

It should also be noted that a number of participants from each group quit 
before the survey loaded. We found during the study that the survey had 
difficulty handling a large number of simultaneous participants. 
A number of experimental group participants also quit during the survey or 
before attempting to recall their password. This may have been due to fatigue; 
as noted in Section \ref{sec:procedure}, participation in the experimental group
took much longer than control group participation did. More information about 
the number of participants at each phase is shown in Figure \ref{fig:retention}.

\section{Password Cracking}
\label{sec:cracking}

During our study, we researched how an adversary would attempt to crack SPARTAN 
passwords offline. None of the open-source password cracking tools
we tried were able to crack SPARTAN passwords off-the-shelf, for two reasons.
First, SPARTAN passwords are simply too long for the password crackers. These 
tools were developed to work on relatively short linear passwords, and the 
source code would have to be edited to accept SPARTAN passwords. Second, they
are not designed to expect the sparseness of SPARTAN passwords. Since each empty
cell was replaced with a space in the password, even if a user used a dictionary
word as his password, the password would be hashed with additional spaces 
present. 
The password cracking tool would need to be updated to compute all of the 
possible permutations of SPARTAN, such as starting in each location in the grid,
being typed in each direction, etc. 

If SPARTAN were deployed in an organization, attackers would readily update
password cracking tools to work with the differently-formatted 
SPARTAN passwords. This tasking was outside the scope of this project and would
make for interesting follow-on work. However, 
cracking a SPARTAN password would still require more 
effort from an attacker than cracking a linear one. 
The major costs to attackers who use 
password cracking tools in an offline stolen password file attack are space 
required to store the dictionary and the amount of time needed to calculate and
search through the hashes. Both of these factors are proportional to the number
of hashes to compute and sort through. Memory is relatively cheap, and
a higher number of 
computations can be defeated by more time or higher computing power. However, 
this is true for linear passwords as well. Even if all user-created SPARTAN passwords were 
input horizontally from left to right, this would still increase the dictionary 
and the number of hashes to search through by a factor of 144 for a 12x12 grid.
This increase could still be defeated with time, but its significance is not
negligible. On the other end of the spectrum, if all users created Point
SPARTAN passwords, this would increase the
dictionary and the number of hashes to search through by a factor of 2.78E+21 
for 10 
character passwords in a 12x12 grid. The computational increase for this is 
significant and could make a huge impact on security. There are many forms of
SPARTAN passwords between these two bounds, and the comparative overall security
of SPARTAN over linear passwords is discussed in the following sections. 

\section{Discussion}
\label{sec:discussion}

Due to the predictability of user-created passwords, many experts in the field
assert that length may be the most important factor in the security of linear 
passwords \cite{NIST_DRAFT-DAG_2016}\cite{SHAY:COMPLEX_TRADEOFF}. One
curious data point gathered during this study is that most SPARTAN passwords 
comprised fewer characters than linear passwords. The median length of a
SPARTAN password created during this study comprised 9 characters, while
the median linear password length was 11 characters.  However, SPARTAN password
locations varied, and this added location factor increases the overall security 
of SPARTAN passwords over that of the linear passwords. There 
are many ways to estimate the security of a user-created password; a few of them
are discussed here for comparison.

\begin{figure}
  \centering
\begin{minipage}[b]{0.23\textwidth}
  \includegraphics[scale=.3]{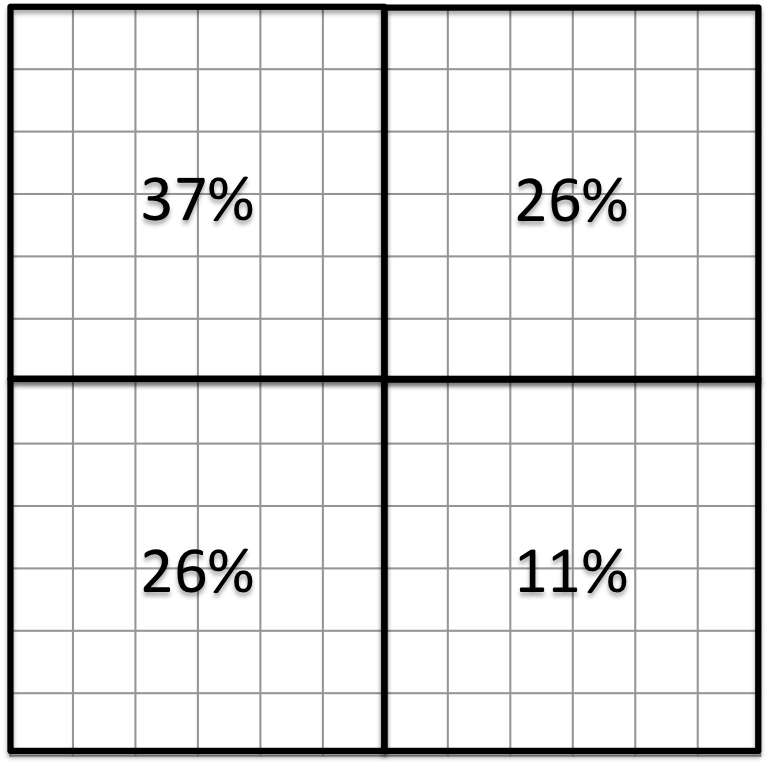}
  \end{minipage}
\hfill
\begin{minipage}[b]{0.23\textwidth}
  \includegraphics[scale=.3]{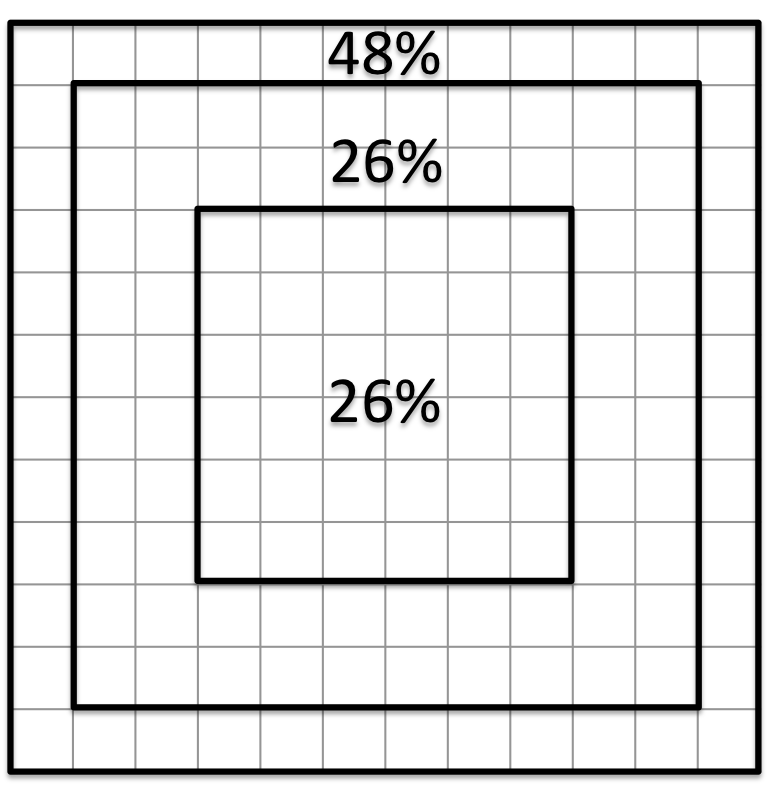}
  \end{minipage}
  \caption{Percent of SPARTAN Straight Line and Snake passwords created 
    beginning in each quadrant of 
    the grid and percent of SPARTAN Straight Line and Snake passwords beginning 
    at the edges versus the 
    center of the grid. N=27}
\label{fig:grids}
\end{figure}

One drawback to the design of our study is that we only gathered the end result 
of password creation. Therefore, while we have the locations of all SPARTAN
passwords, we must make some assumptions about which were the starting and 
ending points of the passwords. Using the mechanics of the prototype and 
previous work in the field demonstrating that users tend to create graphical 
passwords that create a top-down, left-right pattern \cite{passtiles}, we made 
reasonable assumptions about starting location for 
each Straight Line and Snake password created and developed Figure 
\ref{fig:grids}, 
illustrating the occurance of starting locations in each section of the grid.
While users often favored the first quadrant or the edges of the grid as a 
starting location for their password, these areas did not gain a majority of 
users, and many created more complex passwords with less predictable 
starting points.

\subsection{Predicting the Next Cell}

Password security is commonly measured in terms of entropy. Previous work in the
field has tested the validity of this and other measures in relation to 
actual password security against realistic attacks
\cite{ENTROPY_2}\cite{ENTROPY_3}\cite{ENTROPY_1}. We find the 
calculation of entropy added by the use of SPARTAN to be an interesting point 
of discussion for this research. Shannon's work in which he estimated the 
entropy of a string of English text has been used in password security research 
for decades \cite{SHANNON-THE_ORIGINAL}. The 2013 NIST Electronic 
Authentication Guideline builds on Shannon's work and offers similar metrics 
for the entropy of passwords \cite{NIST-EAG_2013}. In their analysis, the first 
character of a password is worth 4 bits of entropy. The following seven 
characters are worth 2 bits each, while characters 9 through 20 are each worth 
1.5 bits. Each character after the twentieth is worth 1 bit each. Using these 
metrics, the median-length linear password created during this study was worth 
23 bits of entropy and the median SPARTAN password text was
worth 20 bits of entropy. 

\begin{figure*}
  \centering
  \includegraphics[scale=.55]{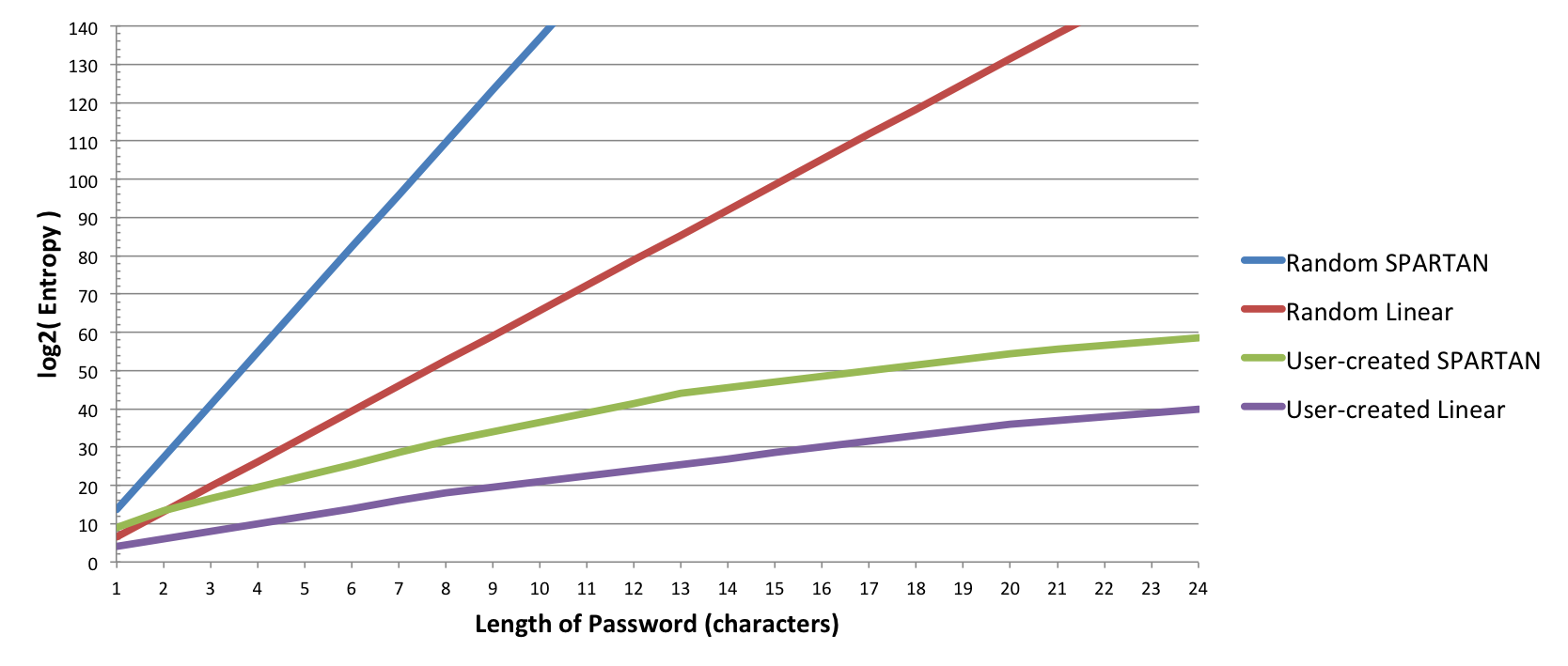}
\caption{Bits of entropy for lengths of user-generated linear and SPARTAN 
passwords and randomly-generated linear and SPARTAN passwords.}
\label{fig:shannon}
\end{figure*}

However, this does not take into account the added security provided by the 
variability
of the location of SPARTAN passwords in the grid. Following a similar method,
we created a formula for the effective entropy of each successive cell used in
a SPARTAN password. Using the data collected from the SPARTAN passwords 
created during this study, we estimated 5 bits of entropy for the first cell 
chosen by the user. The entropy decreases to 2.5 bits for the second cell, due 
to the increased likelihood that the user selects a cell adjacent to the first. 
Cells 3 through 12 are each
worth 1 bit of entropy, and any cell after the twelfth does not provide any 
additional entropy, due to the likelihood that it will follow a pattern 
shown in the previous cells. This cell entropy can be added to the entropy
of the text of the password itself. Using these calculations, Figure
\ref{fig:shannon} shows the comparable entropy for human-created linear and 
SPARTAN passwords of varying lengths. As the graph shows, a 6-character random
linear password has 40 bits of entropy, which is approximately equivalent to an
11-character user-created SPARTAN password, or a 24-character user-created 
linear password. A randomly-generated SPARTAN password reaches 40 bits of 
entropy after just three characters. Thus, using SPARTAN cuts the number of 
characters the user
has to type and remember in half, while maintaining an equivalent amount of 
security in terms of entropy.

\subsection{Dictionary Size}

It is difficult to quantify exactly how bits of entropy correlate to increased 
security. Entropy can also be described as a function of the dictionary size, or
space (S) and the likelihood (L) that a given password will be in the stated
space, as in Equation 1 below \cite{AUTH_TEXT}.
\begin{equation}
Entropy = log_2( S / 2L )
\end{equation}
Looking at the problem from this angle, one can consider the
number of passwords that an attacker might recover using a given dictionary, as
described in Section \ref{sec:cracking}. An 
attacker will try to go after the weakest passwords first when constructing his
dictionary. Assume the adversary has created a dictionary of 5000 possible 
passwords 
and assume he is able to crack all linear passwords comprising a maximum of 
10 characters with this dictionary. Under these assumptions, he would recover 
43\% of the linear passwords created in our study.

Using the same dictionary, but adapting it to look for SPARTAN passwords that 
are up to 10 characters and that start at the top left corner of the 
grid and continue horizontally, 
the dictionary would have the same number of entries, but the adversary would 
only recover
5\% of the passwords created with SPARTAN. Although a greater number of SPARTAN
passwords were 10 characters or fewer, only 5\% of the SPARTAN passwords 
created were both 10 characters or fewer and placed horizontally in the top left
corner. Thus, with the same size dictionary 
(therefore the same number of comparisons to perform, the same amount of time 
spent cracking), the attacker would only recover 5\% of the SPARTAN passwords 
created in this study compared to 43\% of the linear passwords. 

\begin{figure*}
  \centering
  \includegraphics[scale=.50]{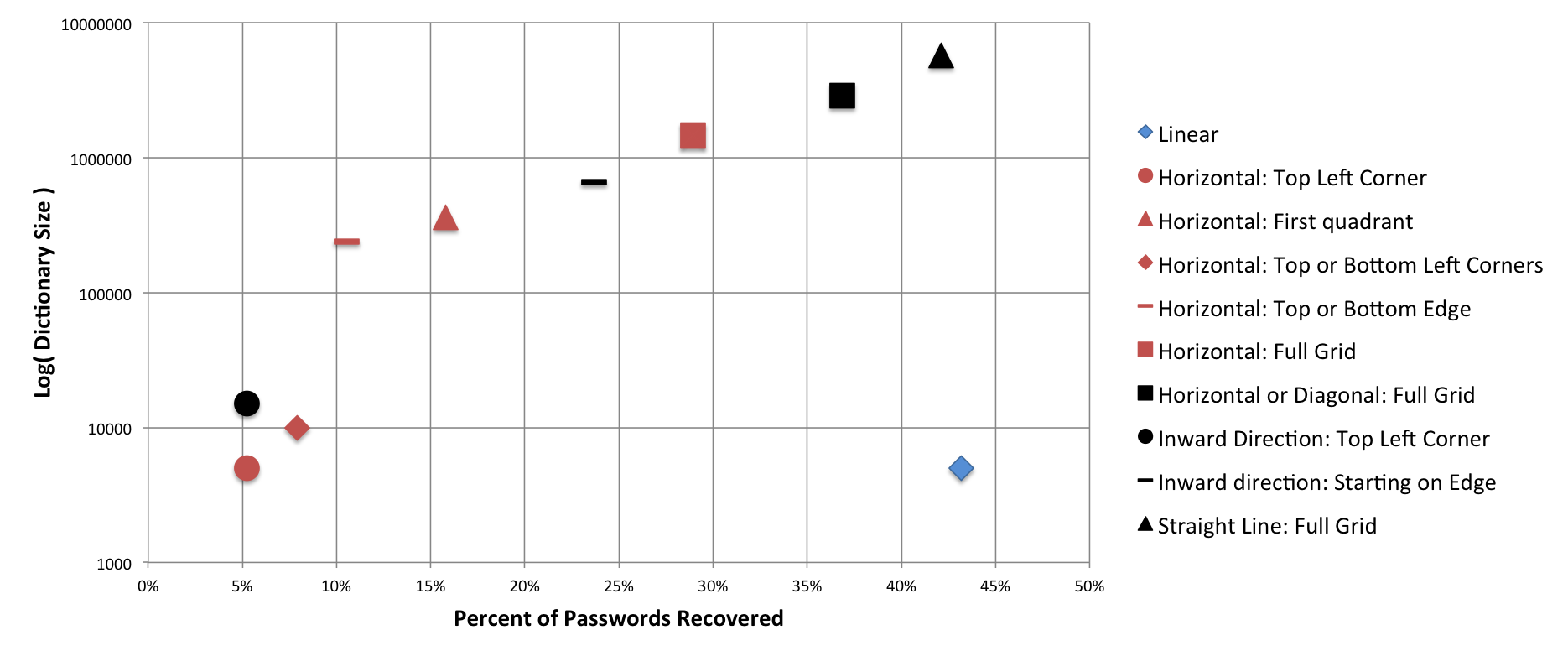}
  \caption{Tradeoff between dictionary size and percent of passwords recovered
for various methods of attack.}
  \label{fig:tradeoff}
\end{figure*}

The attacker could increase his dictionary to include all Straight Line 
horizontal passwords of 10 characters or fewer placed anywhere in the grid. 
By doing so, his dictionary would now be able to recover 29\% of the SPARTAN
passwords created in this study, but his dictionary would necessarily increase
by a factor of 288 for a 12x12 grid (passwords could start anywhere in the
grid, traveling left or right). This trend
continues for other common SPARTAN password placements. The attacker could 
alter his dictionary to target a certain subset of SPARTAN passwords, but
this comes 
with the tradeoff of a larger dictionary, thus more memory and time required.
Figure \ref{fig:tradeoff} displays the calculated tradeoff between dictionary
size for the more common variations of SPARTAN passwords and the percent of 
passwords recovered.

\section{Conclusion and Future Work}
\label{sec:conclusion}

During this user study, we created a SPARTAN prototype, 
gathered empirical data on the security and 
usability of this method, and analyzed SPARTAN password security. 
The 
prototype interface, while sufficiently usable for the purposes of this study, 
would benefit from focused improvement to maximize usability. The usability 
concerns voiced by users were largely against the particular implementation of 
SPARTAN created for this study, and participants were open to new methods of 
authentication. Many users voiced opinions showing that they are overburdened 
with the current state of password requirements employed in their various online
accounts, and they showed that they are interested in trying new forms of 
authentication.

The user-created SPARTAN passwords in this study, while generally shorter than 
the linear passwords created by participants, varied in placement 
throughout the grid and demonstrate that SPARTAN is a promising method for 
shorter, more secure passwords. 
In addition, current password cracking tools are not capable of breaking SPARTAN
passwords without modification, and SPARTAN would necessitate the
use of larger dictionaries than those employed for cracking linear passwords
to recover the same number of user-created passwords.

There is still much work that can be done to evaluate the security and usability
of SPARTAN passwords. First, we would like to address some of the usability 
concerns mentioned by participants in this study. In future deployments, we 
would like to employ a default location and direction in the grid in order to 
ease usability. In addition, we would like to introduce a simpler d-pad 
implementation for the directional arrows instead of superimposed arrows on the 
grid. We would also like to look into mobile adoption, as we can see usability 
on mobile as an important feature going forward.

We gathered 38 SPARTAN and 44 linear passwords 
during this study. In the future, we would like to gather more data on 
user-created SPARTAN passwords to ensure that the trends we 
saw in this study hold with a larger sample size. 
In addition, we would like to conduct a 
study to gain insight into the optimal grid size and layout to balance 
the tradeoffs among usability, memorability, and security. The decreased 
security resulting from the use of a smaller grid (10x10 or 8x8, for example) 
could perhaps be tolerated if increased usability can be demonstrated.
Finally, we would like to conduct a study to gather data on the usability of 
SPARTAN over time, while participants use it
to authenticate periodically over the course of a few months. This 
study would gain data on the memorability of SPARTAN and its usability over 
time, and could incorporate activities related to SPARTAN password changes and 
recovery for analysis.

\bibliographystyle{IEEEtran}
\bibliography{references}  
\end{document}